\documentclass[preprintnumbers,nofootinbib,amsmath,amssymb,twocolumn]{revtex4}
\usepackage{amsfonts}
\usepackage{amssymb}
\usepackage{graphicx}    % needed for including graphics e.g. EPS, PS
\usepackage{amsfonts}
\usepackage{amsmath}
\usepackage{braket}
\usepackage{caption}
\usepackage{rotating}
\usepackage{verbatim}
\usepackage{multirow}

\begin{document}

%\preprint{ArXiv:13XX.XXXX} 
\preprint{Imperial/TP/2015/CC/1} \vskip 0.2in

\title{The bispectrum of single-field inflationary trajectories with $c_{s} \neq 1$.}
%\title{Sound-Speed Non-Gaussianity: The Bispectrum}

\author{Jonathan~S.~Horner}
\affiliation{Theoretical Physics, Blackett Laboratory, Imperial College, London, SW7 2AZ, UK}
\author{Carlo~R.~Contaldi}
\affiliation{Theoretical Physics, Blackett Laboratory, Imperial College, London, SW7 2AZ, UK}
\date{\today}

\begin{abstract}
The bispectrum of single-field inflationary trajectories in which the speed of sound of the inflationary trajectories $c_s$ is constant but not equal to the speed of light $c=1$ is explored. The trajectories are generated as random realisations of the Hubble Slow-Roll (HSR) hierarchy and the bispectra are calculated using numerical techniques that extend the work of \cite{Horner:2013sea}. This method allows for out-of-slow-roll models with non-trivial time dependence and arbitrarily low $c_s$. The ensembles obtained using this method yield distributions for the shape and scale-dependence of the bispectrum and their relations with the standard inflationary parameters such as scalar spectral tilt $n_s$ and tensor-to-scalar ratio $r$. The distributions demonstrate the squeezed-limit consistency relations for arbitrary single-field inflationary models.
\end{abstract}

\maketitle

\section{Introduction}

Current observations of the universe suggest that its density perturbations, to a good approximation, can be considered as a realisation of  a correlated Gaussian statistic and are very close to but not exactly scale independent \cite{Ade:2013ktc,Ade:2013uln,Ade:2015lrj,Ade:2015ava}. This scale dependence is characterised by the measurement of the scalar spectral index $n_{s} = 0.968 \pm 0.006$ \cite{Ade:2015lrj} which agrees well with the framework of the early universe undergoing a phase of quasi-de Sitter expansion that resulted in correlated, super-horizon scaled curvature perturbations to the background metric. The standard, and the most commonly accepted, explanation for both the origin of the perturbations and the reason for the quasi-de Sitter expansion is the presence of a scalar field known as the inflaton whose potential energy dominates the Hubble equation and whose spatial fluctuations seed the curvature perturbations that later drive all structure formation   \cite{Starobinsky198099,PhysRevD.23.347,PhysRevLett.48.1220,1982PhLB..108..389L,Linde1983177,mukhanov1981quantum,mukhanov1982vacuum,hawking1982development,guth1982fluctuations,Starobinsky1982175,PhysRevD.28.679,Mukhanov:1985rz}. 

One of the main issues facing efforts aimed at understanding the nature and origin of the inflaton is that many classes of different inflationary models predict observables such as $n_s$ and $r$ that are in broad agreement with observations (see for example \cite{Yamaguchi:2011kg,Baumann:2009ni, Lyth:1998xn, Wands:2007bd, Tzirakis:2008qy, Silverstein:2003hf}). With the final analysis of Planck data imminent and the combined Planck-BICEPII/Keck analysis \cite{2015PhRvL.114j1301A} confirming that $r$ was in fact {\sl not} detected in the BICEPII data \cite{Ade:2014xna} this situation may become the {\sl status quo} for the foreseeable future. This will be the case unless tensor modes, in the form of $r\ne 0$, are detected by the next generation of sub-orbital Cosmic Microwave Background (CMB) experiments, or, non-Gaussianity is measured. In the former case, discernment between different inflationary models may also require the measurement of the spectral tilt of tensor modes $n_t$ which is challenging due to the cosmic variance effect on the largest scales where the tensor mode signal is clearest.

A detection of non-Gaussianity, in the form of a non-zero bispectrum \cite{2004PhR...402..103B, 2009astro2010S.158K} or un-connected contributions to higher order moments, may then provide the key to uncovering the origin of the inflaton. Non-Gaussianity is necessarily present in the universe since general relativity is a non-linear theory and even if the inflation were driven by a single, free, scalar field it would still interact with gravity giving rise to a non-zero bispectrum. In general, the non-Gaussianity of less standard models of inflation, particularly ones that predict low tensor contributions with $r\to 0$, tends to be large and potentially measurable in the near future.

The bispectrum is the third-order moment of the curvature perturbation in Fourier space and is expected to be the easiest non-Gaussian signal to measure as it is both the lowest order component in the perturbation and has no Gaussian counterpart.  Observational bounds are often quoted in terms of the scale-free amplitude $f_{\mathrm{NL}}$ \cite{Komatsu:2001rj}, a dimensionless quantity which is typically of order the Slow-Roll (SR) parameter $\epsilon \sim 10^{-2}$ for simple inflationary models \cite{Stewart:1993bc,Maldacena:2002vr}. For more complicated models, it is possible to generate a larger $f_{\mathrm{NL}}$ while maintaining $n_{s} \approx 1$ and much effort has been spent constructing such models in the hope that a large non-Gaussianity is detected (see \cite{2004PhR...402..103B,Tzirakis:2008qy,Noller:2011hd,Seery:2005gb,Silverstein:2003hf,Wands:2007bd} for some examples).

Within the context of single field models, there are a couple of
possibilities. One is to break the slow-roll approximation temporarily
by introducing a feature \cite{Chen:2006xjb,Chen:2008wn}, such as a
bump, in the inflaton potential $V(\phi)$. A second is to use a
non-canonical kinetic term for the scalar field \cite{Tzirakis:2008qy,Seery:2005gb,Noller:2011hd,Ribeiro:2012ar}. This involves adding extra derivatives $\partial_{\mu}\phi$ as interactions for the field. One physical consequence of this is that the scalar perturbations typically propagate at a new sound speed $c_{s} < 1$ and it is these models that will be considering in this work.

In this work, for simplicity, we restrict ourselves to the case of a constant $c_{s} \neq 1$, reserving arbitrary time-dependent sound speeds for future work. We calculate the bispectrum of these models numerically, allowing for high values of $c_{s}^{-2} - 1$ and combine this with a Monte Carlo approach for sampling inflationary models. We analyse in detail the exact scale and shape dependence of such models, verifying our results by demonstrating the squeezed-limit consistency relation for very small sound speeds and large SR parameters. 

This {\sl paper} is organised as follows; In Section \ref{HJ_traj} we
summarise the framework and parameters required for the calculation of
the bispectrum and briefly discuss the Monte Carlo generation of
inflationary trajectories using the Hamilton-Jacobi formalism
discussed in more detail in~\cite{Horner:2013sea}. In Section
\ref{PS_calc} we give an overview of the numerical calculation of the
power spectrum before proceeding to the calculation of the bispectrum
in Section~\ref{fnl_calc}. We summarise our results and consistency
checks in Section~\ref{results} before finally concluding in
Section~\ref{conclusion}.

\begin{figure*}[t]
  \begin{center}
    \begin{tabular}{cc}
     \makebox[8.5cm][c]{
    \includegraphics[width=8.5cm,trim=0cm 0cm 0cm 0cm,clip]{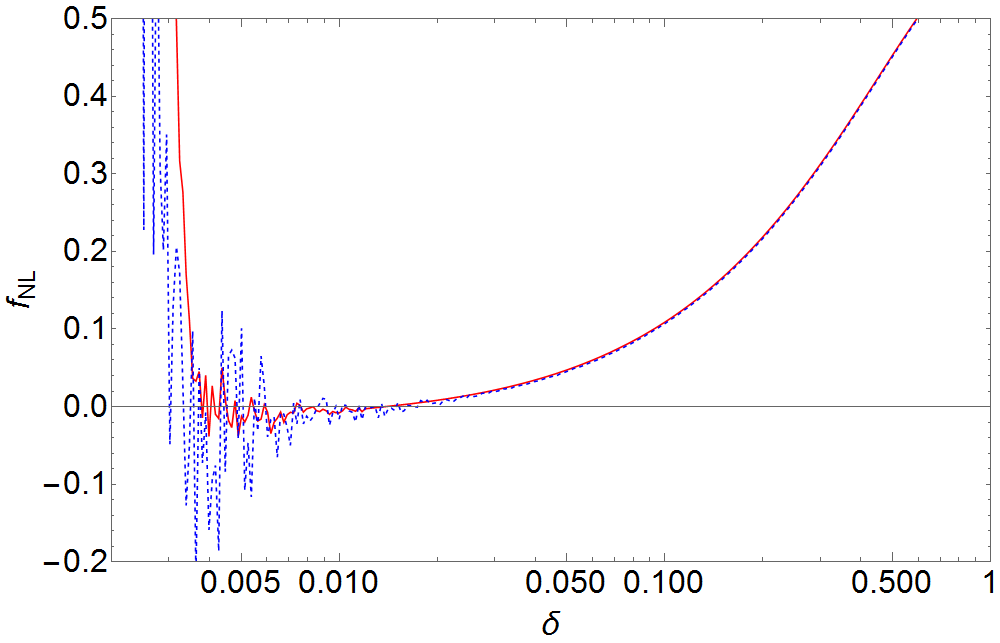}}&
  \makebox[8.5cm][c]{\includegraphics[width=8.5cm,trim=0cm 0cm 0cm
    0cm,clip]{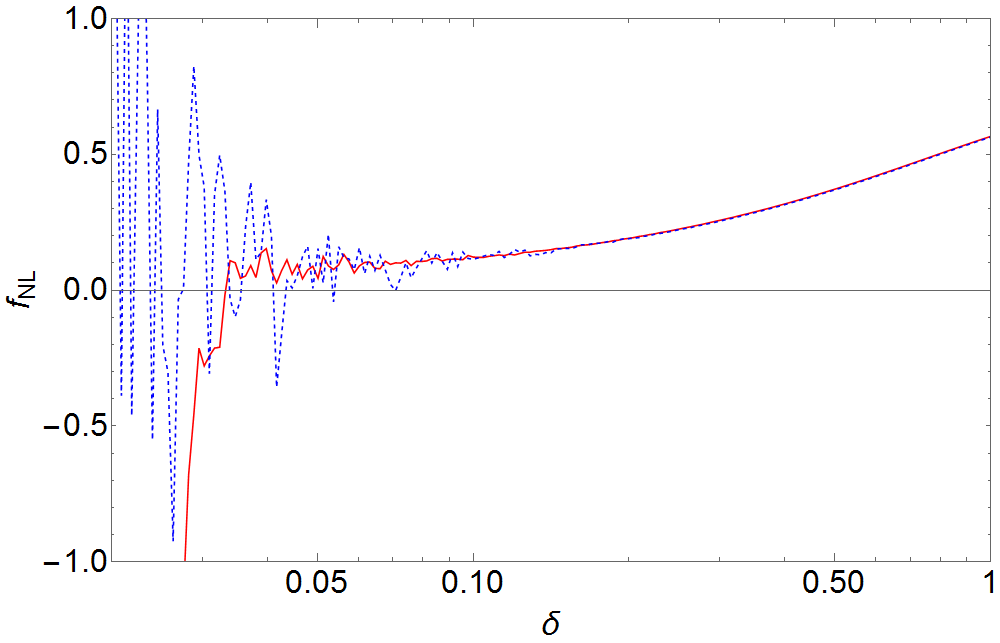} }
  \end{tabular}
  \caption{Dependence of $f_{\mathrm{NL}}$ on the damping factor $\delta$ when $n = 1$ for squeezed (left) and folded (right) configurations. For this trajectory $c_{s}^{-2} = 3$. The red-solid and blue-dashed lines show $k_{min} = 10^{-5} \left(\text{Mpc}\right)^{-1}$ and $k_{max} = 10^{-2} \left(\text{Mpc}\right)^{-1}$ respectively. For large $\delta$ the damping factor is too large affecting the horizon crossing behaviour and the oscillations provide no contribution, producing a smooth curve. For small $\delta$ the oscillations are not sufficiently suppressed producing noise. Ideally $f_{\mathrm{NL}}$ should converge with decreasing $\delta$ in some sense to its true value before the noise begins to dominate. There is an indication of this in the right panel at $\delta \sim 0.1$. Unfortunately for the squeezed limit, the amplitude of $f_{\mathrm{NL}}$ is too small relative to the noise to extract any reasonable result. To make matters worse, depending on the shape the optimum $\delta$ changes by an order of magnitude. Also note noise begins at larger $\delta$ for larger $k$.}
    \label{fig:supp_n_1}
  \end{center}
\end{figure*}

\section{Monte-Carlo approach to sampling trajectories}\label{HJ_traj}

This Hamilton-Jacobi (HJ) formalism \cite{PhysRevD.42.3936,Adshead:2008vn,Liddle:1994dx,Kinney:1997ne}, and its role in numerical inflation was discussed at length in \cite{Horner:2013sea} and we refer the reader to that work for an extended discussion. Here we summarise the method.  In the HJ formalism the dynamics of an inflating cosmology can be captured entirely by considering the Hubble parameter, $H(\phi)$  as a function of the inflaton field value $\phi$ and by considering a hierarchy of Hubble Slow-Roll (HSR) parameters defining the hierarchy if derivatives of $H$ with respect to $\phi$.

We extend this formalism by introducing an arbitrary, but constant sound speed $c_{s} \neq 1$. Following \cite{Garriga:1999vw, Seery:2005gb, Noller:2011hd} we consider actions of the form
\begin{eqnarray}\label{action}
S &=& \int \mathrm{d}^{4}x \sqrt{-g}\,\mathcal{L}\,,\\
\mathcal{L} &=& \frac{M^{2}_{pl}}{2}R + P(X, \phi)\,,\\
X &=& \frac{1}{2}g^{\mu\nu}\partial_{\mu}\phi\partial_{\nu}\phi\,,
\end{eqnarray}
where $M_{pl}$ is the Planck mass, $R$ is the Ricci scalar, and $g^{\mu\nu}$ is the inverse space-time metric. 
The  Lagrangian density $\mathcal{L}$ in the action above describes a perfect fluid with pressure $P(X, \phi)$ and energy density $\rho = 2XP_{X} - P$ where $P_{X} = \partial P / \partial X$. The speed of sound, $c_{s}$, is defined as
\begin{equation}
c_{s}^{2} = \frac{P_{X}}{\rho_{X}} = \frac{P_{X}}{P_{X} + 2XP_{XX}}\,.
\end{equation}
For constant $c_{s}$ this can be treated as a differential equation for $P(X,\phi)$. Using the initial condition $P(X, \phi) = X - V(\phi)$ when $c_{s} = 1$ one obtains
\begin{equation}
P(X, \phi) = \frac{2 c_{s}^{2}}{1 + c_{s}^{2}} X^{\frac{1}{2}(1 + \frac{1}{c_{s}^{2}})} - V(\phi)\,.
\end{equation}

The equation of motion for $\phi$ differs from the canonical case so the original definitions of the HSR parameters in the HJ formalism should be altered accordingly. However, one can still define $e-$foldings $N$, the Hubble rate $H(t)$ and its \emph{time} derivatives independently of the dynamics of the inflation. That is
\begin{eqnarray}\label{SR1}
a(N) &=& e^{N}\,,\\
H(N) &=& \frac{\dot{a}}{a} = \frac{\mathrm{d}N}{\mathrm{d}t}\,,\\
\epsilon(N) &=& - \frac{\mathrm{d}\ln H}{\mathrm{d}N}\,,
\end{eqnarray}
where $a$ is the scale factor and overdots denote differentiation with respect to cosmic time $t$. The HSR parameters can now be defined so that they correspond to the HJ formalism HSR parameters in the limit where $c_{s} = 1$
\begin{equation}\label{SR2}
\frac{\mathrm{d}^{l}\lambda}{\mathrm{d}N} = \left[l\epsilon + (1 - l)\eta\right]\,\,^{l}\lambda - \,\,^{l+1}\lambda\,,
\end{equation}
where $^{1}\lambda = \eta$, $^{2}\lambda = \xi$. 

The values of $^{l}\lambda$ at the end of inflation at $N = N_{tot}$ can be drawn randomly to sample the distribution of consistent inflationary trajectories as described in \cite{Horner:2013sea}. The sound speed will not affect the time dependence of these parameters so it will not play an explicit role in the sampling of trajectories . In practice the random sampling is achieved by drawing the following set of parameters with uniform distributions (flat prior) in the intervals
\begin{eqnarray}\label{prior}
^{l}\lambda = [-1,1]xe^{-sl}\,\\
N_{tot} = [60,80] + \ln A\,,
\end{eqnarray} 
where $l>0$. In addition since we draw samples at the end of inflation we fix the value of the $l=0$ HSR parameter $^{0}\lambda\equiv \epsilon(N_{tot}) =  1$.

In (\ref{prior}), $x$ and $s$ are parameters that specify the scaling of the uniform prior range with $l$ and can be used to investigate the dependence of our final results on the assumed priors. The random sampling of $N_{tot}$ represents the uncertainty in the total duration of the post-inflationary reheating phase and the constant $A$ is related to the normalisation of $H$ which will be discussed shortly. Formally one would need to evolve an infinite number of $^{l}\lambda$ parameters to sample the space of all possible $H(N)$ functions. In practice  this is not possible and one must truncate the series at some finite order $L_{\max}$. We define $L_{\max}$ such that $L_{\max}$ HSR parameters includes $\epsilon(N)$ e.g. $L_{\max} = 2$ corresponds to $\epsilon(N)$ and $\eta(N)$ with all other $^{l}\lambda = 0$ identically. Once random values of $^{l}\lambda$ have been drawn the entire inflationary trajectory can be obtained by integrating the background equations of motion sufficiently far back in the past to cover the required number of $e$-foldings given by $N_{tot}$.

\section{Computational method}\label{comp_method}

The calculation of the bispectrum relies on the same basic building blocks as the calculation of the primordial power spectrum. In addition the bispectrum is often compared to the spectral tilt of the power spectrum and the squeezed limit consistency condition is a valuable tool for checking the numerical method. We therefore give a brief review the calculation of the power spectrum as the first step in the numerical calculation of the bispectrum.

\subsection{Computation of the power spectrum}\label{PS_calc}

We choose a gauge where the inflaton perturbation $\delta\phi(t, \textbf{x}) = 0$ and the spatial metric is given by $g_{ij} = a^{2}(t)e^{2\zeta(t, \textbf{x})}\delta_{ij}$. This defines the comoving curvature perturbation $\zeta(t, \textbf{x})$. The primordial power spectrum of the curvature perturbation is then
\begin{equation}
\langle\zeta_{k_{1}}\zeta^{*}_{k_{2}}\rangle = (2\pi)^{3}\delta^{(3)}(\textbf{k}_{1} + \textbf{k}_{2})P_{\zeta}(k_{1})\,,
\end{equation}
where $\textbf{{k}}$ is the wavevector of the Fourier mode and $k = |\textbf{k}|$. These modes satisfy the Mukhanov-Sasaki equation \cite{Mukhanov:1985rz,Sasaki:1986hm} which, with our choice of variables becomes
\begin{equation}\label{mukh}
\frac{\mathrm{d}^{2}\zeta_{k}}{\mathrm{d}N^{2}} + (3 + \epsilon - 2\eta)\frac{\mathrm{d}\zeta_{k}}{\mathrm{d}N} + \left(\frac{c_{s}k}{aH}\right)^{2}\zeta_{k} = 0\,.
\end{equation}

To obtain the power spectrum we simply require the freeze-out value of $\zeta_{k}$ when the mode crosses the sound-horizon, i.e.
\begin{equation}
P_{\zeta}(k) = \left.|\zeta|^{2}\right|_{c_{s}k \ll aH}\,
\end{equation}
notice that for theories where the speed of sound and light are not equivalent the horizon set by the speed of sound is the relevant scale beyond which freeze-out occurs.

We apply the usual Bunch-Davies initial conditions \cite{Bunch:1978yq} when the mode is deep inside the sound-horizon
\begin{equation}\label{BunchDavies}
\zeta_{k} \to \frac{1}{2M_{pl}a}\sqrt{\frac{c_{s}}{k\epsilon}}e^{-ic_{s}k\tau}\,,
\end{equation}
where $\tau$ is  conformal time defined through $\mathrm{d}N/\mathrm{d}\tau = aH$.

We impose initial conditions (\ref{BunchDavies}) at different
$e$-folds for each mode $k$. This ensures all modes are sufficiently
deep inside the sound-horizon at the start of the forward integration
of (\ref{mukh}). The starting $e$-folds, $N_{k}$, for mode with
wavenumber $k$ is set by requiring that $c_{s}k = A\,aH(N_{k})$ where
$A \gg 1$. In practice this means that the integration is started at
successively later times as $k$ increases. This avoids unnecessary
computational steps at smaller scales.

\begin{figure*}[t]
  \begin{center}
    \begin{tabular}{cc}
     \makebox[8.5cm][c]{
    \includegraphics[width=8.5cm,trim=0cm 0cm 0cm
    0cm,clip]{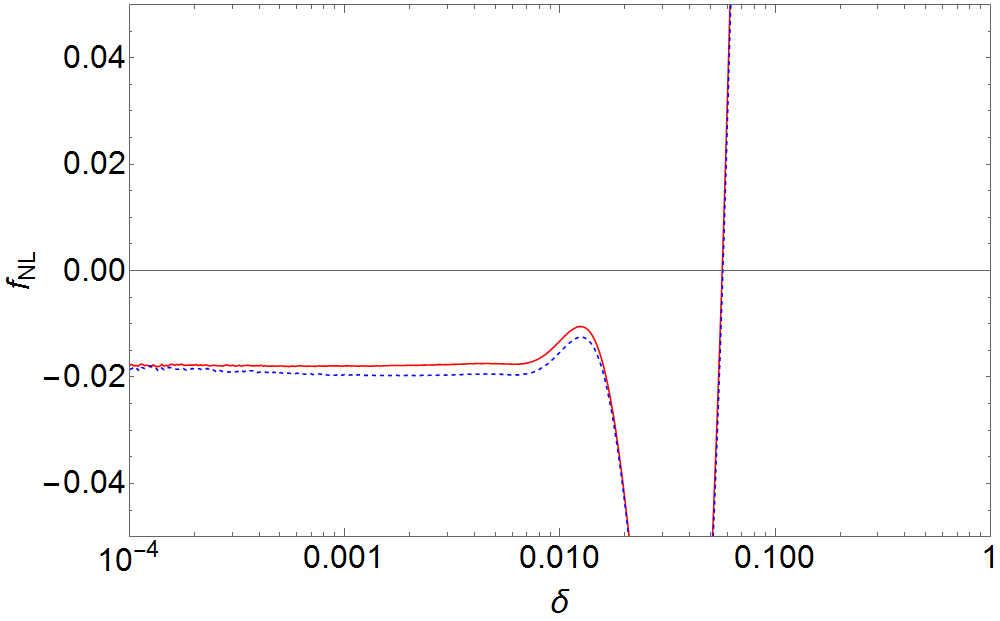}}&
  \makebox[8.5cm][c]{\includegraphics[width=8.5cm,trim=0cm 0cm 0cm
    0cm,clip]{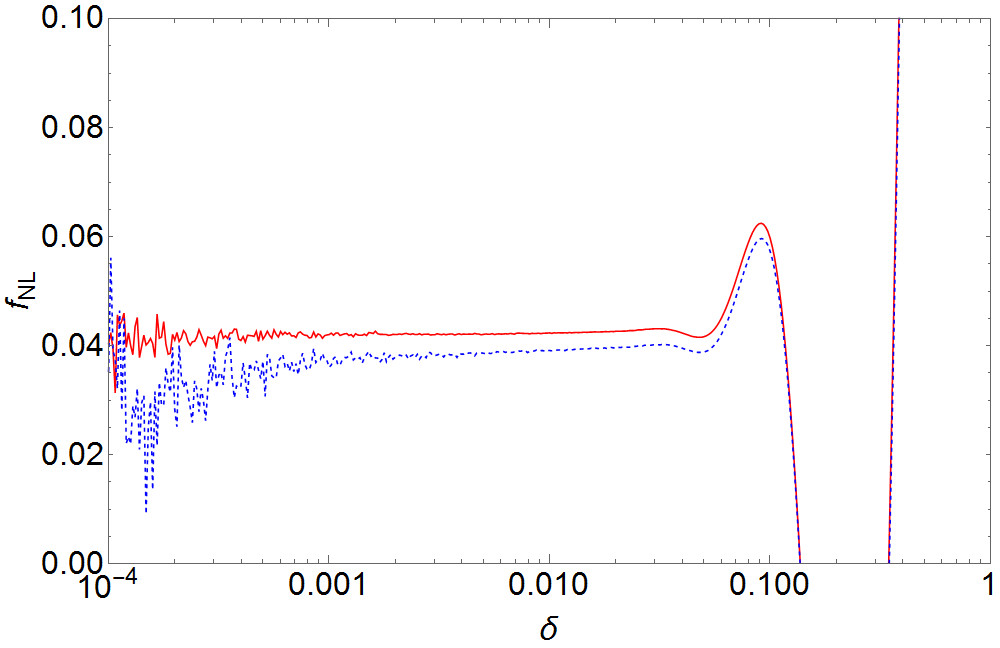} }
  \end{tabular}
  \caption{Dependence of $f_{\mathrm{NL}}$ on the damping factor $\delta$ when $n = 3$ for squeezed (left) and folded (right) configurations. For this trajectory $c_{s}^{-2} = 3$. The red-solid and blue-dashed lines show $k_{min} = 10^{-5} \left(\text{Mpc}\right)^{-1}$ and $k_{max} = 10^{-2} \left(\text{Mpc}\right)^{-1}$ respectively. By choosing $n > 1$ the suppression is weighted more towards the early time oscillations and less on the horizon crossing time. Practically this pushes the noise back to very small values of $\delta$ allowing $f_{\mathrm{NL}}$ to converge to its true value. The acceptable range of $\delta$ is also much wider solving the shape dependence problem. One could choose $n \gg 1$ but most of the time $c_{s}k/aH > 1$. Therefore to prevent damping at horizon crossing $\delta$ must be reduced to compensate and this method can only be pushed so far. In practice we found $n = 3$ to be sufficient. Note for large $k$ the noise still arises at larger $\delta$. }
    \label{fig:supp_n_3}
  \end{center}
\end{figure*}

Assuming $H(N)$ varies slowly enough, each mode will evolve for roughly $\ln A$ $e$-folds before they cross the sound-horizon and freeze out. The earliest mode of interest to freeze out will be $k_{\min}$ so we choose $N_{k_{\min}} = 0$, i.e. $N = 0$ is defined such that $c_{s}k_{\min} = A\,aH(N = 0)$ and we then apply (\ref{BunchDavies}) to this mode. This means the $k_{\min}$ mode will cross the sound-horizon at $N_{c}\approx \ln A$ and we can then use the standard analytical result relating $H$ to the amplitude of the power spectrum to normalise $H$. In practice, during the backwards integration of the HSR parameters, we apply a normalisation condition on $H$ such that 
\begin{equation}
H(N_{c}) = 2\pi \sqrt{2\, c_{s}\epsilon(N_{c})}A_{s}M_{pl}\,,
\end{equation}
where $A_{s}$ is conventional the normalisation of the dimensionless primordial curvature power spectrum. In the usual power law convention for the form of the power spectrum $A_{s}$ is employed as
\begin{equation}
k^{3}P_{\zeta}(k) = A^{2}_{s}\left(\frac{k}{k_{min}}\right)^{n_{s} - 1}\,.
\end{equation}

A similar procedure can be carried out for the calculation of the gravitational wave spectrum which is unaffected by $c_{s}$. The analogues of  (\ref{mukh}) and (\ref{BunchDavies}) are are identical to the standard case with $c_s=1$
\begin{eqnarray}\label{mukh_tensor}
\frac{\mathrm{d}^{2}h_{k}}{\mathrm{d}N^{2}} + (3 - \epsilon)\frac{\mathrm{d}h_{k}}{\mathrm{d}N} + \left(\frac{k}{aH}\right)^{2}h_{k} = 0\,,\\
h_{k} \to \frac{1}{M_{pl}}\frac{e^{-ik\tau}}{a\sqrt{2k}}\,.&&
\end{eqnarray}

A complication that arises due to the sound and light horizon not
being the same is that scalar and tensor modes freeze out at different
times so one must be sure that the Bunch-Davies conditions are applied
when both modes are sufficiently deep inside their respective
horizons. In principle the power spectrum must converge in the limit
$A \to \infty$ therefore the answer should not depend on whether the
Bunch-Davies conditions are applied earlier to one mode with respect
to another as long as both modes are sufficiently deep inside their
respective horizons. In practice this means nothing needs to be
changed. If $c_{s}k = A\,aH$ then we know $k \geq A\,aH$ as $c_{s}
\leq 1$ so the tensor mode is even deeper inside its respective
horizon than the scalar mode is. The only concern is a penalty to
computational efficiency as the modes become highly oscillatory when
they deep within their horizon.

With all the integration constants fixed, the full set of differential
equations (\ref{SR1})-(\ref{SR2}), (\ref{mukh}) and
(\ref{mukh_tensor}) can be integrated until both the scalar and tensor
modes are well outside the sound and light horizons respectively. This
requirement can be parametrised by a constant $B \ll 1$. Following the
same argument, if $k = B\, aH$, we have $c_{s}k \leq B\,aH$ as $c_{s}
\leq 1$. In summary we integrate the mode equations from a time such
that $c_{s}k = A\,aH$ until $k = B\,aH$ with $A \gg 1$ and $B \ll
1$. When calculating the bispectrum (for isosceles triangles) we have
a third horizon to consider for the squeezed/folded mode. Similar
arguments can be made and one should take care to ensure all relevant
modes exit their horizons and satisfy the relevant initial
conditions. we have We found the bispectrum to converge when $A \sim
400$ and $B \sim 1/100$. Higher values of $A$ significantly increased
the computation time due to the oscillatory nature of the mode
functions while providing no real benefit. Smaller values of $B$ did
not affect the accuracy or computation time.

\begin{figure*}[t]
  \begin{center}
    \begin{tabular}{cc}
     \makebox[8.5cm][c]{
    \includegraphics[width=8.5cm,trim=0cm 0cm 0cm
    0cm,clip]{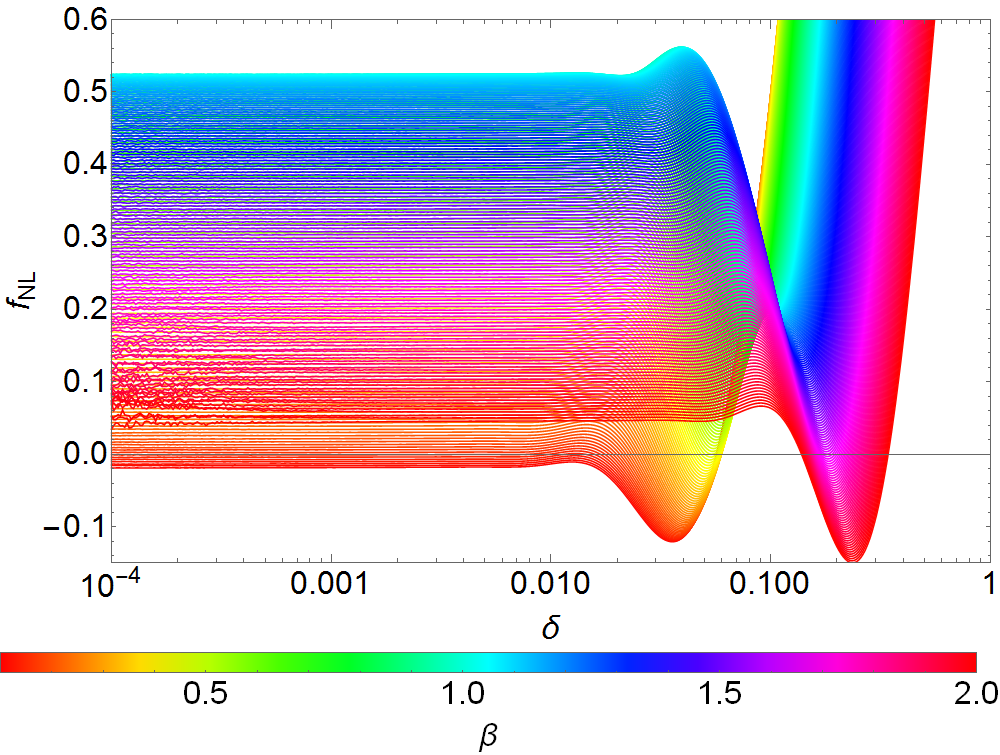}}&
  \makebox[8.5cm][c]{\includegraphics[width=8.5cm,trim=0cm 0cm 0cm
    0cm,clip]{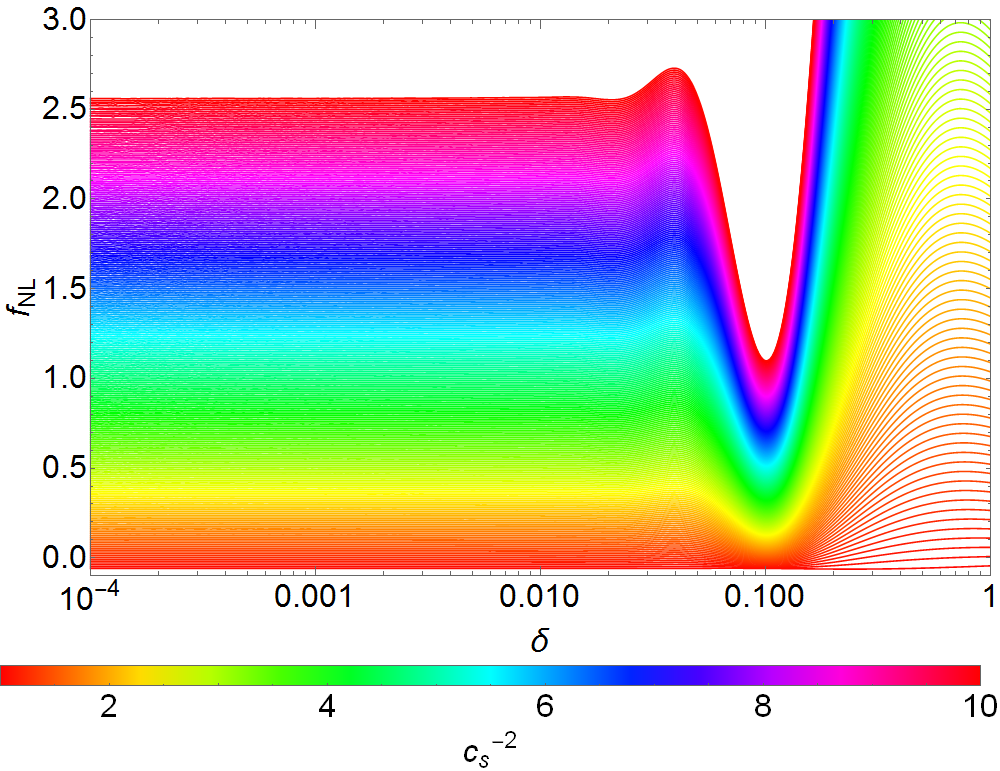} }
  \end{tabular}
  \caption{Dependence of $f_{\mathrm{NL}}$ on $\delta$ on shape and sound speed for the smallest scale  $k = k_{\text{max}} = 10^{-2} \left(\text{Mpc}\right)^{-1}$. The optimum delta occurs when the relevant curve has converged. The left panel shows how the $\delta$ dependence varies for each shape evaluated at $c_{s}^{-2} = 3$. There is a mild shape dependence in the optimum $\delta$ where squeezed triangles require smaller $\delta$ values. As a consequence, noise from folded configurations occurs at larger values of $\delta$ so the optimum $\delta$ must lie between these two cases. The right panel shows how the $\delta$ dependence varies with sound speed dependence evaluated in the equilateral limit. There is remarkably little dependence on $c_{s}$ even at very small sound speeds.}
    \label{fig:supp_n_3_equil}
  \end{center}
\end{figure*}

With the scalar and tensor power spectra in hand, the observables $n_{s}$ and $r$ can be calculated directly following their definitions, either as a function of scale $k$ or at a specific ``pivot'' scale $k_\star$ for comparison with conventional models
\begin{eqnarray}
n_{s}(k_{\star}) & = & 1 + \left.\frac{\mathrm{d}\ln \left(k^{3}P_{\zeta}(k)\right)}{\mathrm{d}\ln k}\right|_{k = k_{\star}}\,,\\
r(k_{\star})& = & 8\frac{P_{h}(k_{\star})}{P_{\zeta}(k_{\star})}\,,
\end{eqnarray}
where the factor of 8 in the definition of $r$ arise from the definition of the tensor perturbations and from the fact that two independent polarisations contribute to the total power.

\subsection{Computation of the bispectrum}\label{fnl_calc}

The bispectrum of $\zeta$ is the simplest, lowest-order moment, where we expect to see deviations from a pure Gaussian statistic. It corresponds to a tree-level three-point vertex for an interacting quantum field and will be the most dominant form of non-Gaussianity as higher order moments are expected to be suppressed by higher order terms in both the HSR parameters and level of curvature perturbations with  $A_{s}^{1/2}\sim 10^{-5}$. In the isotropic limit it reduces to a function of three variables, the magnitudes of the wavevectors $\mathbf{k}_{1}$,  $\mathbf{k}_{2}$, and $\mathbf{k}_{3}$ making up the allowed, closed triangles in Fourier space
\begin{equation}
\langle\zeta_{k_{1}}\zeta_{k_{2}}\zeta_{k_{3}}\rangle = (2\pi)^{3}\delta^{(3)}(\mathbf{k}_{1}+\mathbf{k}_{2}+\mathbf{k}_{3})B(k_{1}, k_{2}, k_{3})\,,
\end{equation}
where the delta function imposes the closed triangle condition due to isotropy. We define the reduced, dimensionless, scale and shape dependent bispectrum as
\begin{eqnarray}\label{fnl_def}
f_{\mathrm{NL}}(k_{1}, k_{2}, k_{3}) = 5B(k_{1}, k_{2}, k_{3})/&&\nonumber\\
6\left(\left|\zeta_{k_{1}}\right|^{2}\left|\zeta_{k_{2}}\right|^{2} + \left|\zeta_{k_{1}}\right|^{2}\left|\zeta_{k_{3}}\right|^{2} + \left|\zeta_{k_{2}}\right|^{2}\left|\zeta_{k_{3}}\right|^{2}\right)\,,&&
\end{eqnarray}
This is different to the usual $f_{\mathrm{NL}}$, scale free, amplitude for the bispectrum quoted in the literature \cite{Komatsu:2001rj}.

\begin{figure*}[t]
  \begin{center}
    \begin{tabular}{cc}
     \makebox[8.5cm][c]{
    \includegraphics[width=8.5cm,trim=0cm 0cm 0cm
    0cm,clip]{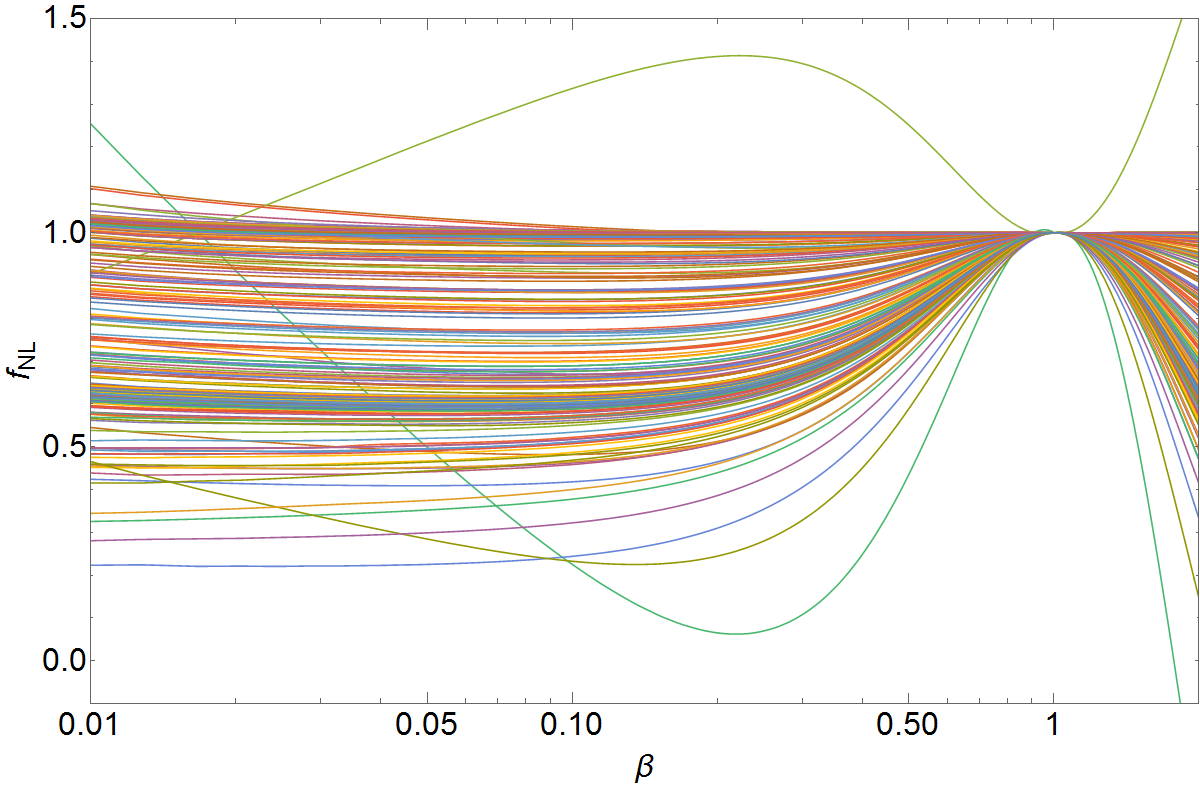}}&
  \makebox[8.5cm][c]{\includegraphics[width=8.5cm,trim=0cm 0cm 0cm
    0cm,clip]{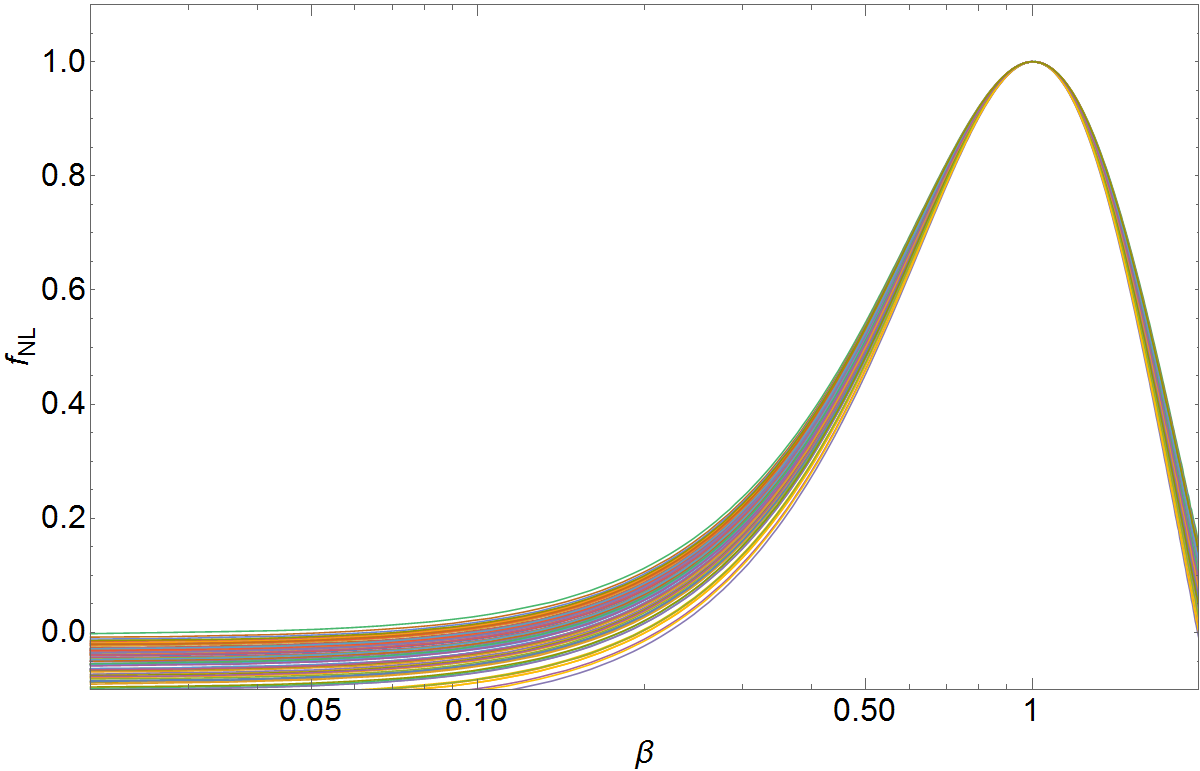} }
  \end{tabular}
  \caption{Shape dependence of $f_{\mathrm{NL}}$ for several trajectories evaluated at $c_{s}^{-2} = 1$ (left) and $c_{s}^{-2} = 3$ (right). All values of $f_{\mathrm{NL}}$ are normalised to their value at $\beta = 1$. Single field inflation models generically peak in the equilateral limit but because they must follow the consistency relation in the limit $\beta \to 0$ their $\beta$ dependence is much sharper.}
    \label{fig:shape}
  \end{center}
\end{figure*}

The weighting introduced in the definition of $f_{\rm NL}$
(\ref{fnl_def}) is known as the ``local'' weighting. Other definitions
are used in the literature depending on the expected shape dependence
of the signal. When observational constraints are obtained from data,
such as with Planck \cite{Ade:2015ava} the various choices of
weighting are used to define limits on {\sl different} types of
$f_{\rm NL}$. These include  equilateral and orthogonal weightings. The
limits reported in \cite{Ade:2015ava} are $f_{\rm NL}^{\rm local} = 0.8
\pm 5.0$ , $f_{\rm NL}^{\rm equil} = -4\pm 43$, $f_{\rm NL}^{\rm ortho}= -26\pm 21$.

The most dominant contribution to the bispectrum comes from (\ref{action}) expanded to third order in $\zeta$. Following \cite{Seery:2005gb, Noller:2011hd} the third-order action for single field inflation with a constant sound speed $c_{s}$ is
\begin{widetext}
\begin{eqnarray}\label{S_3}
S_{3} = M^{2}_{pl}\int \mathrm{d}^{4}x\left[\frac{2a^{3}\epsilon}{3Hc_{s}^{2}}\left(\frac{1}{c_{s}^{2}} - 1\right)\dot{\zeta}^{3} + \frac{a^{3}\epsilon}{c_{s}^{2}}\left(\frac{2\eta - \epsilon}{c_{s}^{2}} + 3\left(1 - \frac{1}{c_{s}^{2}}\right)\right)\zeta\dot{\zeta}^{2} + \frac{a\epsilon}{c_{s}^{2}}(\epsilon + 1 - c_{s}^{2})\zeta(\partial\zeta)^{2}\right.\\
\left. + \frac{a\epsilon}{c_{s}^{2}}(\eta - \epsilon)\zeta^{2}\partial^{2}\zeta - \frac{2a^{3}\epsilon^{2}}{c_{s}^{4}}\left(1 - \frac{\epsilon}{4}\right)\dot{\zeta}\partial_{i}\zeta\partial_{i}\partial^{-2}\dot{\zeta} + \frac{a^{3}\epsilon^{3}}{4c_{s}^{4}}\partial^{2}\zeta\partial_{i}\partial^{-2}\dot{\zeta}\partial_{i}\partial^{-2}\dot{\zeta}\right]\,.
\end{eqnarray}
\end{widetext}
Section~III.B of \cite{Horner:2013sea} discussed why the action is written in the form (\ref{S_3}) in order to deal with apparent divergences and we refer the reader to that work for further detail. A $c_s\ne 1$, and indeed, an arbitrary time-dependant $c_{s}$ provides no further complications in dealing with the third-order action.

The "In-In formalism" \cite{Maldacena:2002vr, Noller:2011hd, Seery:2005gb} is used to calculate the bispectrum and ultimately $f_{\mathrm{NL}}$. Using (\ref{S_3}) to define an interaction Hamiltonian and treating $\zeta(t, \textbf{x})$ as a scalar field with canonical commutation relations, the bispectrum can be reduced to a single integral over $N$.
\begin{equation}\label{bispectrum}
B(k_{1}, k_{2}, k_{3}) = \Im\left[\zeta^{\star}_{1}\zeta^{\star}_{2}\zeta^{\star}_{3}\int^{N_{1}}_{N_{0}}\!\!\!\mathrm{d}N\,\,Z(N)\right]\,.
\end{equation}
Here $\Im[z]$ denotes the imaginary part of the imaginary number $z$. $N_{0}$ and $N_{1}$ represent times when the largest and smallest scales are sufficiently deep inside and far outside the sound-horizon respectively, using the same $A$ and $B$ parameters as described above. $Z(N)$ implicitly depends on the shape and scale of the triangle but the function arguments have been omitted for brevity.

We now specialise to the case where $k_{1} = k_{2} = k$ and $k_{3} = \beta k$ where $0 <\beta \leq 2$. This simple parametrisation covers many cases of interest. The squeezed, equilateral, and folded limits correspond to $\beta = 0, 1$ and 2 respectively. $Z(N)$ then takes on the following form:
\begin{eqnarray}\label{Z}
Z(N) & = & \!\!\frac{5Ha^{3}\epsilon}{3c_{s}^{2}}\left(f_{1}\zeta'^{2}\zeta'_{\beta} + f_{2}\zeta^{2}\zeta_{\beta} + f_{3}\zeta\zeta'\zeta'_{\beta} + f_{4}\zeta'^{2}\zeta_{\beta}\right)\,,\nonumber\\
f_{1}& = & 4u\,,\nonumber\\
f_{2}& = & \left(2 + \beta^{2}\right)\left(\frac{c_{s}k}{aH}\right)^{2}\left(u + \frac{1}{c_{s}^{2}}(2\eta - 3\epsilon)\right)\,,\\
f_{3}& = & 12u - \frac{2}{c_{s}^{2}}\left(4\eta + (1 - \beta^{2})\epsilon +\left(\frac{\beta^{2}}{4} - 1\right)\epsilon^{2}\right)\,,\nonumber\\
f_{4}& = & 6u - \frac{1}{c_{s}^{2}}\left(4\eta + 2(\beta^{2} - 1)\epsilon +\left(\frac{\beta^{2}}{4} - 1\right)\beta^{2}\epsilon^{2}\right)\,,\nonumber
\end{eqnarray}
where $\zeta = \zeta_{k}$, $\zeta' = \mathrm{d}\zeta/\mathrm{d}N$, $\zeta_{\beta} = \zeta_{\beta k}$ and $u = 1 - c_{s}^{-2}$. At early times in the limit $A \to \infty$, $|Z(N)| \to \infty$. However we deform the integration contour by a small, imaginary component $i\delta$ so that the oscillations arising from (\ref{BunchDavies}) become exponentially suppressed. This is the usual choice of contour one makes when calculating interacting correlation functions. In this limit (\ref{BunchDavies}) becomes
\begin{equation}\label{BD_rotated}
\zeta_{k} \to \lim_{\delta \to 0} \frac{1}{2M_{pl}a}\sqrt{\frac{c_{s}}{k\epsilon}}e^{-ic_{s}k(1 + i\delta)\tau}\,,
\end{equation}
as $\tau \to -\infty$ and the integral converges at very early times.

\subsubsection{Regulating the integral}

\begin{figure*}[t]
  \begin{center}
    \begin{tabular}{cc}
     \makebox[8.5cm][c]{
    \includegraphics[width=8.5cm,trim=0cm 0cm 0cm
    0cm,clip]{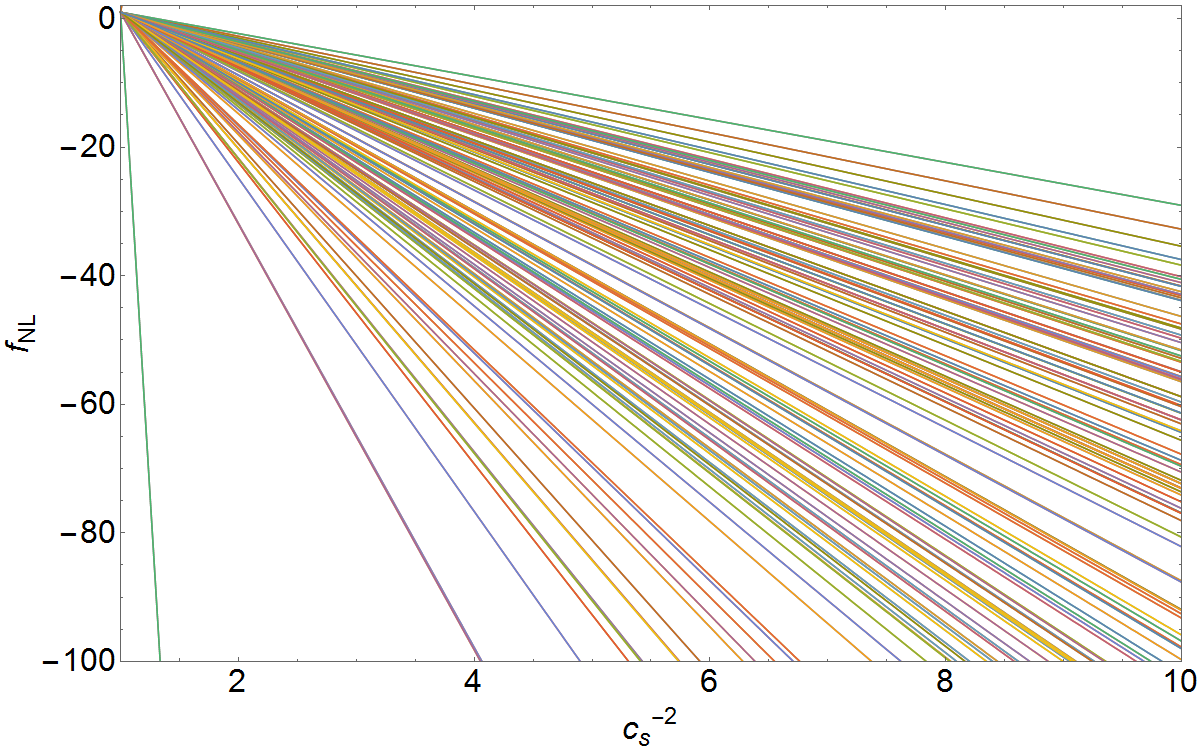}}&
  \makebox[8.5cm][c]{\includegraphics[width=8.5cm,trim=0cm 0cm 0cm
    0cm,clip]{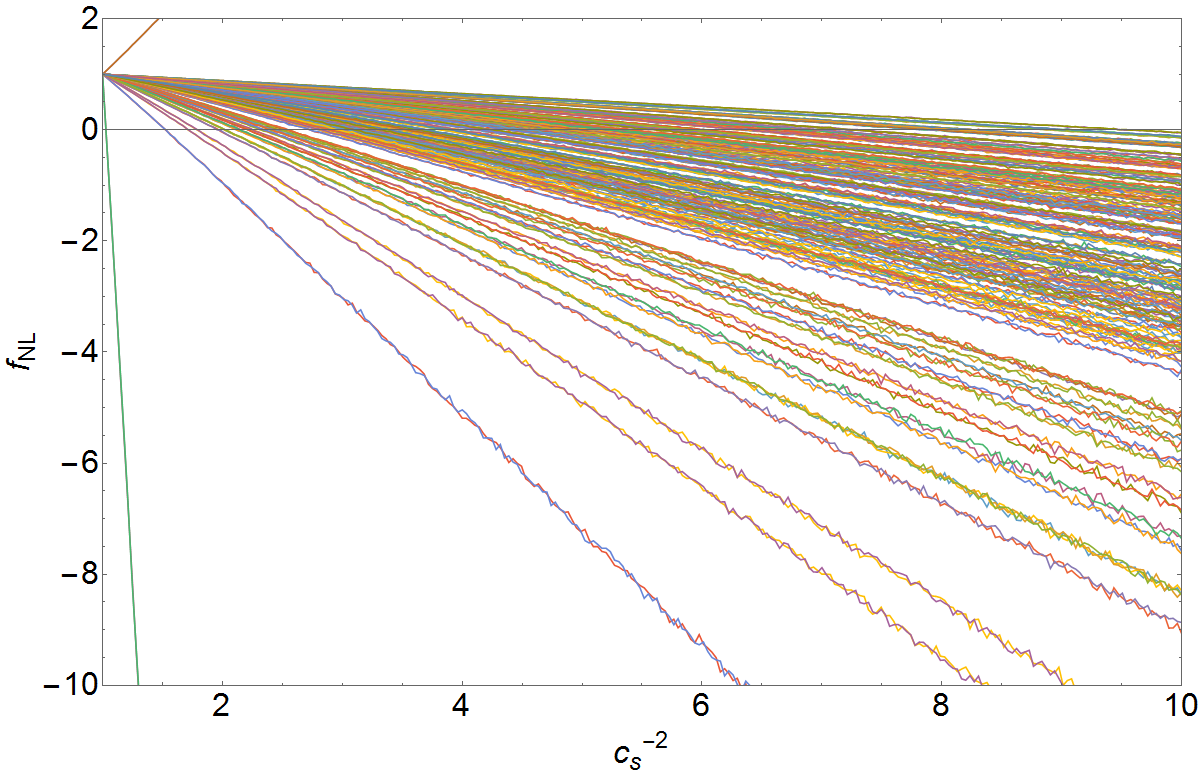} }
  \end{tabular}
  \caption{$c_{s}$ dependence of $f_{\mathrm{NL}}$ for several trajectories evaluated at $\beta = 1$ (left) and $\beta = 0.1$ (right). All values of $f_{\mathrm{NL}}$ are normalised to their value at $\beta = 1$. For equilateral triangles the $\beta$ dependence is much stronger. In the squeezed limit the $c_{s}$ dependence becomes much smaller but remains non-zero.}
    \label{fig:sound}
  \end{center}
\end{figure*}

To calculate the bispectrum we integrate  (\ref{bispectrum}) numerically. Analytically, after performing the integral, one could take the limit $\delta \to 0$ to obtain an answer that is well behaved. Unfortunately this is not possible numerically and gives rise to large errors. We cannot integrate over an infinite range in time, i.e. from $A = \infty$, $a(N) = 0$ or $N = -\infty$, so there will always be a sharp integration cutoff at very early times. Because of this sharp cutoff, the oscillations in the integrand result in large fluctuations in the final answer even though they should cancel out if the integration constant is formally extended to $-\infty$.

A solution o this problem is to add an exponential damping factor similarly to the one introduced in (\ref{BD_rotated}). This was the first approach taken by Chen {\it et. al.} in \cite{Chen:2006xjb}. However there are some issues with this method. Firstly the amplitude of the integrals tend to be suppressed resulting in an underestimation of the bispectrum. In addition, the optimal value for the damping factor $\delta$ needs to be fine tuned for each scale considered \cite{Chen:2006xjb}.

An alternative method exists which which does not suffer from these issues. It was first used in \cite{Chen:2008wn} and then expanded on in \cite{Horner:2013sea}. We refer the reader to \cite{Horner:2013sea} for the details.  The method splits the integral into two parts at an arbitrary split point defined by $c_{s}k = X aH$. $X$ needs to be large enough for (\ref{BunchDavies}) to be a good approximation for all  three modes. Some integration by parts is performed then $X$ is chosen to minimise the error on the bispectrum. Unfortunately this method does not work for $c_{s} \neq 1$ because of the new $\zeta'^{3}$ term. The method still prevents the contributions from the oscillations at early times from diverging but the $\zeta'^{3}$ term still introduces a large oscillatory signature to the final integral. We therefore adopted the first method employing an improved exponential damping factor
\begin{equation}
Z(N) \to Z(N)e^{-\delta\left(\frac{c_{s}k}{aH}\right)^{n}}\,,
\end{equation} 
in the numerical integration.

Fig.~\ref{fig:supp_n_1} shows the dependence of $f_{\mathrm{NL}}$ on the suppression factor $\delta$ for $n = 1$ in both the squeezed and folded limits. For this figure, and all the other $\delta$ dependence figures, a random trajectory was taken with $s = 1.5$, $x = 1$ and $L_{\text{max}} = 4$, as defined in (\ref{prior}). We see that if $\delta$ is too small, the early time oscillations are not sufficiently suppressed producing a large amount of noise. This noise is exaggerated for large values of $k$. Secondly, if $\delta$ is too large, the damping factor will interfere with the time dependence around the time of horizon crossing. This is the most dominant contribution to the integral so it will no longer be a good approximation to the bispectrum. For this choice of $n=1$ it is hard to justify an optimal value of $\delta$ where $f_{\mathrm{NL}}$ has converged. 

Another issue is that the optimal $\delta$ depends on the shape of the triangle. Indeed, between the folded and squeezed cases the optimal $\delta$ drops by an order of magnitude. This dependence can be reduced by adjusting the value of $n$. $c_{s}k/aH$ is very large at early times and of order 1 during horizon crossing. Therefore increasing $n$ will give stronger weighting to the damping factor at early times, while interfering less with the horizon crossing time. We found $n = 3$ to give the best results. The $\delta$ dependence for $n = 3$ is shown in Fig.~\ref{fig:supp_n_3}.

Most of the residual noise arises from large $k$ modes, particularly in the folded configuration. In contrast most results calculated in the equilateral configuration are relatively clean. Fig.~\ref{fig:supp_n_3_equil} shows how the $\delta$ dependence varies with shape factor $\beta$ and $c_{s}$ in the equilateral configuration. Fig. \ref{fig:supp_n_3} motivates a choice of $\delta \approx 0.005$ and we use this suppression factor along with $n=3$ for the remainder of our calculations.

\section{Results}\label{results}

\begin{figure*}[h]
  \begin{center}
    \begin{tabular}{cc}
    	\makebox[7.5cm][c]{\includegraphics[width=6.5cm,trim=0cm 0cm 0cm 0cm,clip]{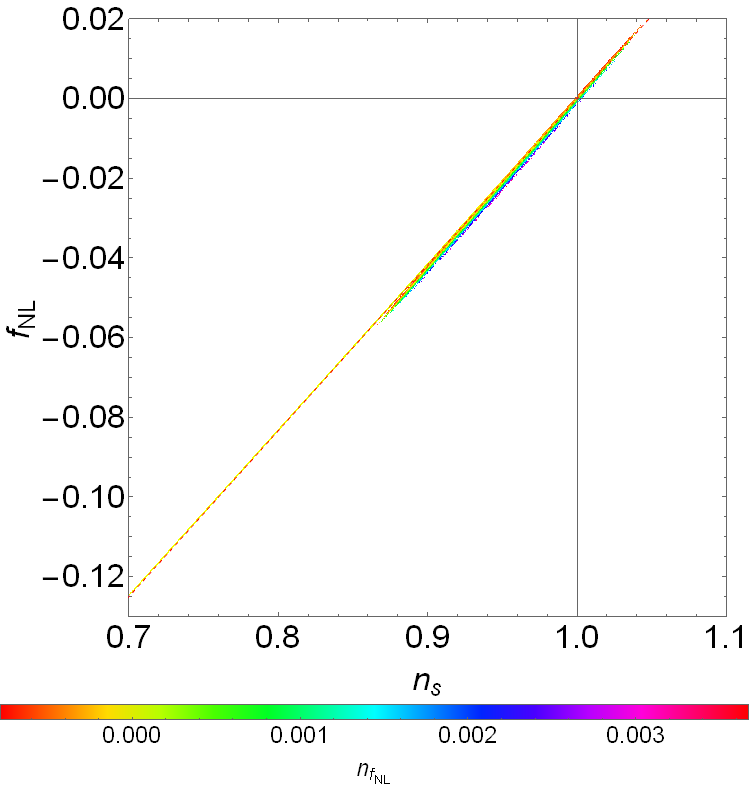}}&
  		\makebox[7.5cm][c]{\includegraphics[width=6.5cm,trim=0cm 0cm 0cm 0cm,clip]{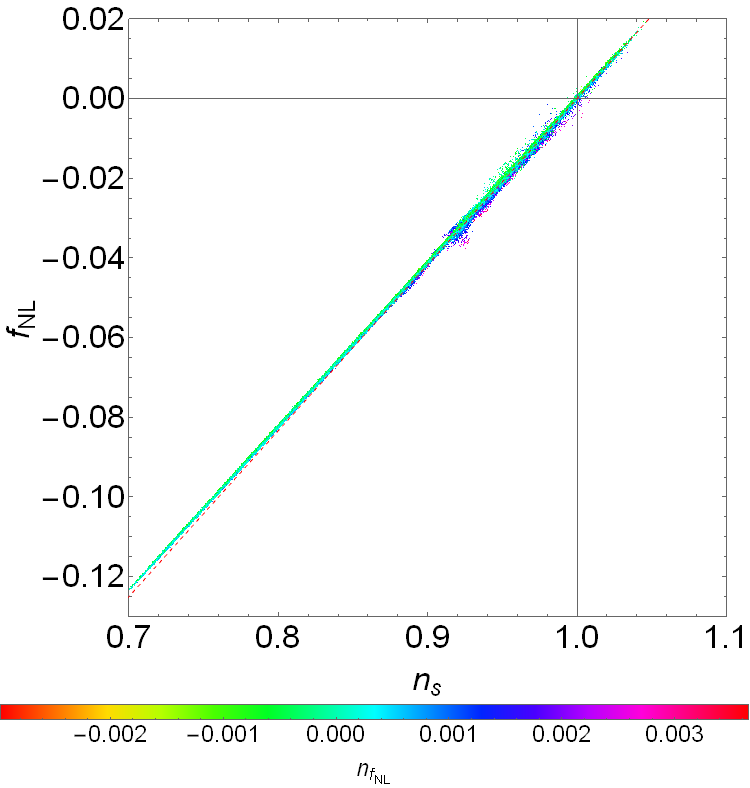}}\\
    	\makebox[7.5cm][c]{\includegraphics[width=6.5cm,trim=0cm 0cm 0cm 0cm,clip]{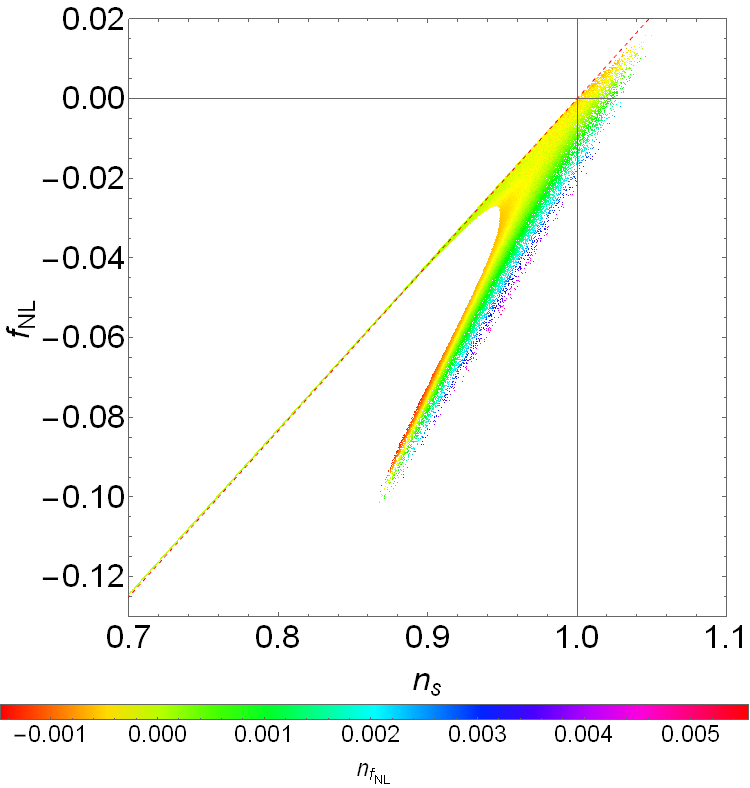}}&
		\makebox[7.5cm][c]{\includegraphics[width=6.5cm,trim=0cm 0cm 0cm 0cm,clip]{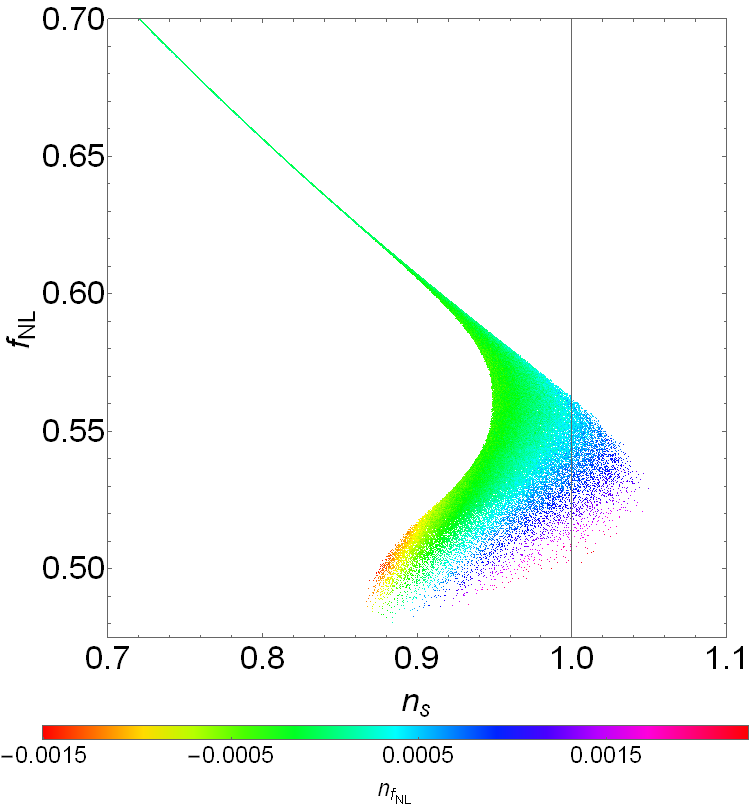}}\\
		\makebox[7.5cm][c]{\includegraphics[width=6.5cm,trim=0cm 0cm 0cm 0cm,clip]{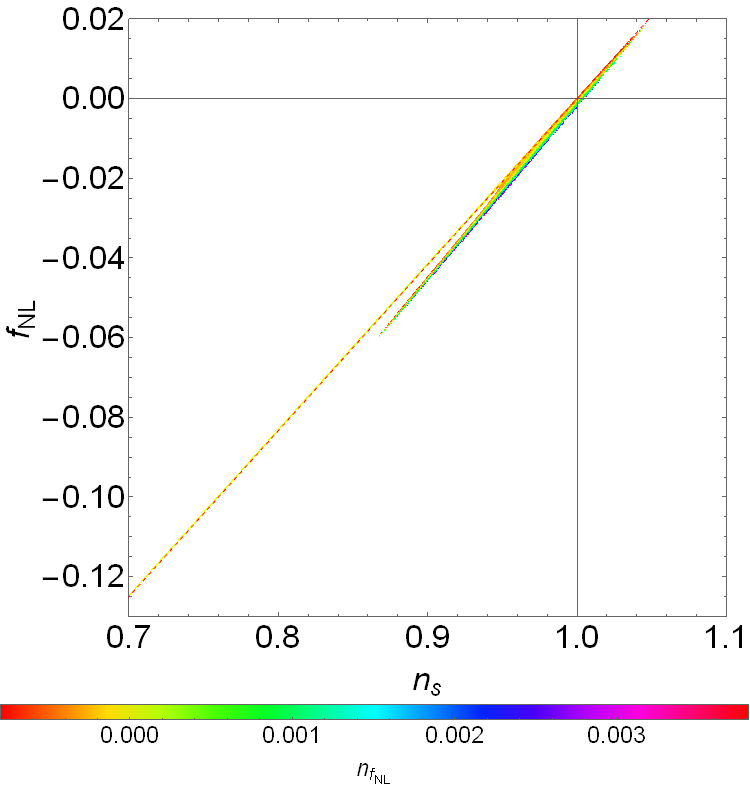}}&
		\makebox[7.5cm][c]{\includegraphics[width=6.5cm,trim=0cm 0cm 0cm 0cm,clip]{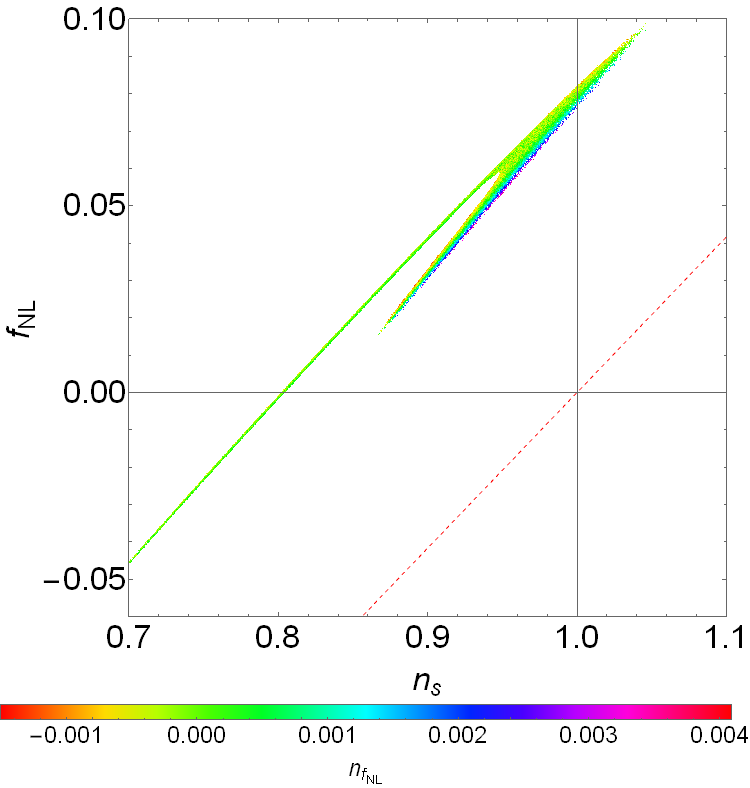}}
  \end{tabular}  \end{center}
  \caption{Monte Carlo plots for $c_{s} = 1$ (left) and $c_{s}^{-2} =
    3$ (right). From top to bottom the shape configurations are evaluated in the squeezed, equilateral and folded limits respectively. The red-dashed line represents the consistency relation $5(n_{s} - 1)/12$. The colour of each trajectory illustrates the scale dependence of the bispectrum, $n_{f_{\mathrm{NL}}}$. For squeezed $c_{s}^{-2} = 3$ (top-right) it was necessary to reduce $\beta = 0.02$ to recover the squeezed limit as opposed to $\beta = 0.1$ for the $c_{s} = 1$ case. This increased computation time by roughly an order of magnitude.}
    \label{fig:MC}
\end{figure*}

One way to test our numerical results for robustness and consistency is  by comparison with the the squeezed limit consistency relation \cite{Maldacena:2002vr,Creminelli:2004yq}. For \emph{any} single field inflation model the following limit must hold 
\begin{align}
\lim_{k_{3} \ll k_{1}, k_{2}}\!\!\! \langle\zeta_{k_{1}}\zeta_{k_{2}}\zeta_{k_{3}}\rangle\to
(2\pi)^{3}\delta^{(3)}(\textbf{k}_{1}+\textbf{k}_{2}+\textbf{k}_{3})\left(n_{s} - 1\right)P_{k_{1}}P_{k_{3}}\,
\end{align}
or in our notation
\begin{equation}
\lim_{\beta \to 0} f_{\mathrm{NL}}\to \frac{5}{12}\left(n_{s} - 1\right)\,.
\end{equation}

It is important to emphasise here that this holds for all single field models independent of the value of $c_{s}$ or the prior we choose for the initial conditions of the background trajectories. However increasing the value of $c_{s}$ or the HSR parameters typically increases the amplitude of $f_{\mathrm{NL}}$ therefore we don't necessarily expect all models to tend to the squeezed limit at the same rate. For example $\beta = 0.1$ might be ``squeezed enough'' for low values of $c_{s}$ but not for higher values. We first analyse the shape and sound speed dependence of the trajectories, elaborating on the consistency relation in section~\ref{MC_plots}. Unless stated otherwise, the trajectories are taken from a prior with $x = 1, s = 1.5$ and $L_{\text{max}} = 4$.

\subsection{Shape dependence}

Fig.~\ref{fig:shape} compares the shape dependence of trajectories evaluated at $k_{\text{min}} = 10^{-5} \left(\text{Mpc}\right)^{-1}$ normalised to their equilateral values. As expected, for trajectories with shape dependence $|f_{\mathrm{NL}}|$ peaks in the equilateral configuration. As $c_{s}$ reduces, the amplitude of $|f_{\text{NL}}|$ typically increases but the trajectories must still obey the squeezed limit consistency relation where $|f_{\text{NL}}| \sim 10^{-2}$. This exaggerates the shape dependence of all the trajectories, even those which appear flat when $c_{s} = 1$.\\
It is worth noting that in the squeezed limit, the shape dependence is curved in comparison to the roughly linear dependence in the folded limit. This is in agreement with \cite{2011JCAP...11..038C} where the authors show that corrections linear in $\beta$ drop out. Any terms linear in $\textbf{k}_{3}$ must contract symmetrically with the remaining two modes. As they have equal magnitudes in opposite directions they will cancel out leaving only quadratic corrections in $k_{3}$. In the folded limit this cancellation does not occur producing the linear dependence shown in Fig.~\ref{fig:shape}.

\subsection{$c_{s}$ dependence}

Fig. \ref{fig:sound} compares the dependence of $f_{\mathrm{NL}}$ on $c_{s}$ for equilateral and squeezed triangles. These values are normalised to their values at $c_{s} = 1$. To a good approximation the dependence is linear in $c_{s}^{-2}$ and much stronger for equilateral triangles. This shows that for fixed $\beta = 0.1$ one can still obtain large $f_{\mathrm{NL}}$ by choosing an arbitrarily small $c_{s}$.  At $c_{s} = 1$, $f_{\mathrm{NL}}$ is typically small and negative so as $c_{s} \to 0$ $f_{\mathrm{NL}}$ becomes large and positive. The close linear dependence on $c_{s}^{-2}$ is not surprising and it clearly arises from the functions $f_{i}$ in (\ref{Z}).

\subsection{Monte Carlo Plots}\label{MC_plots}

The scale dependence is linear to a good approximation and can easily be analysed. To this end we define $n_{f_{\mathrm{NL}}}$ as
\begin{equation}
n_{f_{\mathrm{NL}}}(k_{\star}, \beta) = \left.\frac{\mathrm{d}f_{\mathrm{NL}}(k, \beta)}{\mathrm{d}\ln k}\right|_{k = k_{\star}}\,.
\end{equation}

As discussed in \cite{Chen:2005fe,2009JCAP...12..022S,2010JCAP...02..034B} it is possible to define a scale dependence as long as the shape of the triangle is kept fixed. Our definition is different to the usual definition of $n_{f_{\mathrm{NL}}}$ which is the derivative of $|\ln f_{\mathrm{NL}}|$ and this is simply to avoid difficulties arising when $f_{\mathrm{NL}} \approx 0$. Recall, reducing $c_{s}$ often in induces a sign change as can be seen in Fig.~\ref{fig:sound}.

Fig.~\ref{fig:MC} shows numerous Monte Carlo plots for various sound speeds and shapes. Each plot consists of $2^{18}$ trajectories with their colour representing $n_{f_{\mathrm{NL}}}$. The top two figures show that all trajectories tend towards the squeezed limit consistency relation even for small sounds speeds $c_{s} < 1$. The consistency relation $5(n_{s} - 1)/12$ is shown by the red-dashed line. To reach the consistency relation in the $c_{s}^{-2} = 3$ case, a much smaller $\beta$ was required (and consequently the value of $\delta$ had to be lowered, recall Fig.~\ref{fig:supp_n_3}).

In the equilateral case one can see clearly how a small sound speed deforms the inflationary attractor. For example in the $c_{s} = 1$ case, the consistency relation acts as a firm upper limit for $f_{\mathrm{NL}}$. The deviation from the consistency relation is simply proportional to $\epsilon > 0$ and $f(k) > 0$ defined in \cite{Maldacena:2002vr}. A small $c_{s} < 1$ clearly violates this relation deforming the distribution significantly, resulting in a large positive $f_{\mathrm{NL}}$. In the folded limit, the distribution is reduced back again to be parallel with the consistency relation, although this time with a positive,$c_{s}$ dependent offset.

\begin{figure*}
	 \begin{center}
    \begin{tabular}{cc}
    	\makebox[7.5cm][c]{\includegraphics[width=6.5cm,trim=0cm 0cm 0cm 0cm,clip]{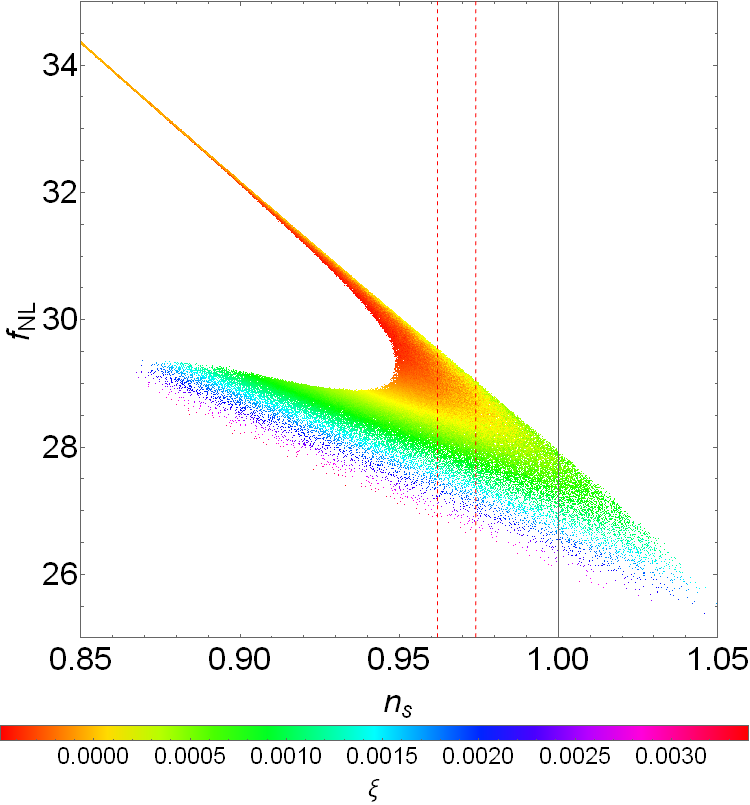}}&
  		\makebox[7.5cm][c]{\includegraphics[width=6.5cm,trim=0cm 0cm 0cm 0cm,clip]{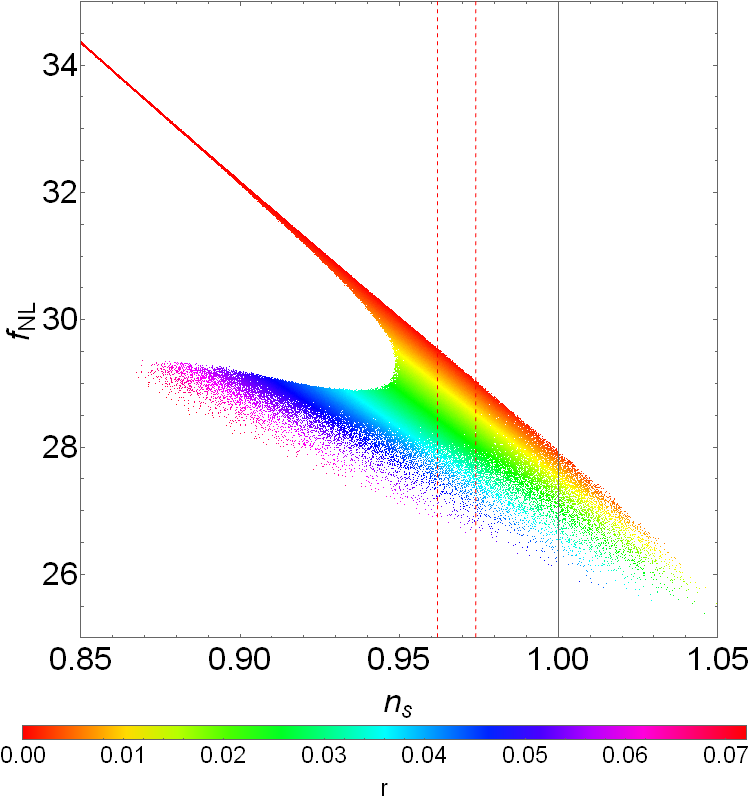}}
  \end{tabular}
  \caption{Monte Carlo plot for a very small sound speed $c_{s}^{-2} =
    100$ evaluated in the equilateral limit. The red-dashed lines
    represent the recent Planck constraints on $n_{s}$ \cite{Ade:2015lrj}. The left panel shows how small variations in $\xi$ can change $f_{\mathrm{NL}}$. The right panel shows the corresponding tensor-to-scalar ratio $r$ for each trajectory.  }
    \label{fig:very_small_cs}
  \end{center}	
\end{figure*}

To illustrate the flexibility of the method
Fig.~\ref{fig:very_small_cs} shows a distribution with $c_{s}^{-2} =
100$ with colour of the trajectories now representing the third slow
roll parameter $\xi=\,^{2}\!\lambda$ evaluated shortly after horizon
crossing and the tensor-to-scalar ratio $r$. The dashed lines
represent the current Planck constraints on $n_{s} = 0.968 \pm 0.006$
\cite{Ade:2015lrj}. Planck also constrains  $f_{\rm NL}^{\rm equil} =
-4\pm 43$ \cite{Ade:2015ava} although it is important to remember that
there is not an exact one-to-one correspondence between our
$f_{\mathrm{NL}}$ calculated here and the one constrained by Planck \cite{Ade:2015ava} due to assumptions on scale-invariance.

For example in power law inflation $\xi = \epsilon^{2}$ and is often assumed to be vanishingly small. However at these sound speeds, one can see that a small variation in $\xi$ can lead to an appreciable change in $f_{\mathrm{NL}}$ even though it is likely to be neglected.

From the right panel in Fig.~\ref{fig:very_small_cs} one can also see
that for small $c_{s}$ tighter constraints on $r$ require larger
$\left|f_{\mathrm{NL}}\right|$. From one perspective this is not
surprising as, to leading order, $r \approx 16\,c_{s}\epsilon$ \cite{Garriga:1999vw} so smaller sounds speeds naturally induce smaller $r$. However one has to remember that the right panel in Fig.~\ref{fig:very_small_cs} shows trajectories for \emph{fixed} $c_{s} = 0.1$. The changes in $f_{\mathrm{NL}}$ and $r$ can \emph{only} be induced by the slow-roll parameters (and $H$). More concretely smaller values of $\epsilon$ are thus expected to produce \emph{more} non-Gaussianity. This is in contrast to the $c_{s} = 1$ case where \emph{larger} values in $\epsilon$ produce more non-Gaussianity. Indeed it is often quoted that $f_{\mathrm{NL}}\sim\epsilon$. From the plots this is fairly easy to explain. Increasing $\epsilon$ always contributes negatively to $f_{\mathrm{NL}}$. It just so happens that at $c_{s} = 1$, $f_{\mathrm{NL}}$ is small and negative so they add constructively. On the other hand reducing $c_{s}$ always contributes positively to $f_{\mathrm{NL}}$ eventually inducing a sign change. As soon as $f_{\mathrm{NL}}$ changes sign, increasing $\epsilon$ reduces the amount of non-Gaussianity.  

\section{Discussion}\label{conclusion}

We have outlined a  full, numerical calculation of the bispectrum  with a particular emphasis on single field models of inflation with non-canonical speed of sound. The calculation is challenging due to the oscillatory nature of the integrands involved which is exacerbated for the case with $c_s\neq 1$ and we have shown how regularising the integrals can lead to stable results with the correct choice of numerical damping terms. The methods explored in this work can be used to investigate the scale and shape dependence of the bispectrum signal produced by an  epoch of inflation.

For convenience we have adopted a more general description of bispectrum signal than that normally quoted in the literature by re-defining a scale and shape dependent $f_{\mathrm{NL}}$, which always tends to $5(n_{s} - 1)/12$ as the shape parameter $\beta \to 0$. For lower values of $c_{s}$, $|f_{\mathrm{NL}}|$ is typically much greater and thus requires much smaller values of $\beta$ to recover the squeezed limit consistency relation.

If future observational surveys of the CMB or large scale structure become accurate enough to constrain any scale dependence of the non-Gaussian signal then our work could be applied to the calculation of accurate model of the bispectrum to be used in likelihood evaluations of the data. This is not currently possible as the strongest limit on non-Gaussianity come from an ad-hoc analysis of Planck CMB maps assuming a scale-independent and fixed shape templates for the bispectrum leading to constraints on a single amplitude parameter. Whilst these results may be consistent with the simplest model of inflation, if a non-zero amplitude for $f_{\mathrm{NL}}$ were ever to be measured, more accurate parametrisations of the non-Gaussianity will be useful to try to gain a better understanding of the nature of the inflaton and its connection with extensions to the standard model of particle physics. This will particularly become a priority if primordial tensor modes are not discovered at levels $r\sim 0.01-0.1$.

\begin{acknowledgements}
 JSH is supported by a STFC studentship.
\end{acknowledgements}

\bibliography{paper}

\end{document}